\documentclass[aps,prl,twocolumn,superscriptaddress,amsmath,amssymb,showpacs]{revtex4}
\usepackage{graphicx}
\usepackage{dcolumn}
\usepackage{bm}
\begin{document}

 \title{NMR evidence for two-step phase-separation in ${\rm
Nd_{1.85}Ce_{0.15}CuO_{4-\delta}}$.}

 \author{O. N. Bakharev}

 \author{I. M. Abu-Shiekah}

 \author{H. B. Brom}

 \affiliation{Kamerlingh Onnes Laboratory, Leiden University, POB 9504,
2300 RA Leiden, The Netherlands}

 \author{A. A. Nugroho}

 \affiliation{Van der Waals-Zeeman Institute, University of Amsterdam,
1018 XE Amsterdam, The Netherlands}

 \altaffiliation{MSC, University of Groningen, Neijenborg 4, 9747 AG
Groningen, The Netherlands}

 \author{I. P. McCulloch}

 \author{J. Zaanen}

 \affiliation{Instituut Lorentz for Theoretical Physics, Leiden
University, POB 9506, 2300 RA Leiden, The Netherlands}

\begin{abstract}
By Cu NMR we studied the spin and charge structure in ${\rm
Nd_{2-x}Ce_{x}CuO_{4-\delta}}$.  For $x=0.15$, starting from a
superconducting sample, the low temperature magnetic order in the
sample reoxygenated under 1 bar oxygen at 900$^0$ C, reveals a
peculiar modulation of the internal field, indicative for a phase
characterized by large charge droplets ('Blob'-phase). By
prolonged reoxygenation at 4 bar the blobs brake up and the spin
structure changes to that of an ordered antiferromagnet (AF). We
conclude that the superconductivity in the n-type systems competes
with a genuine type I Mott-insulating state.
\end{abstract}

\pacs{PACS numbers: 76.60.-k, 74.72.Dn, 75.30.Ds, 75.40.Gb \\May
17 2004 accepted for Phys. Rev. Letters}

\maketitle

At present an important issue in cuprate superconductivity is the
nature of the electronic state(s) competing with the
superconducting state (see \cite{Sachdev03,Kivelson03} and
references therein). In the hole doped cuprates evidence has been
accumulating that at low dopings this competitor is a stripe phase
\cite{Tranquada95}. It is believed that these stripe phases find
their origin in the microscopic incompatibility between the
metallic- and Mott-insulating states \cite{Zaanen89,Emery93}.
Stripes are a-priori not a unique way of resolving this
incompatibility. This was sharply formulated recently in terms of
the analogy with superconductivity \cite{Lee03}; stripes can be
viewed as the analogue of the type II superconductor while a type
I behavior is also imaginable. In the latter the Mott-insulator
and the metal/superconductor are immiscible and one expects a
state characterized by droplets of metal/superconductor in a
Mott-insulating background. The electron doped (n-type)
superconductors are microscopically \cite{Zaanen85} quite
different from the hole doped ones, while also macroscopic
properties seem distinct \cite{Imada98}. So far, no incommensurate
spin fluctuations have been found by neutron scattering in this
system \cite{Yamada99,Vajk02,Mang03}, arguing against stripe
phases. Instead, recently it was reported that field induced {\em
commensurate} antiferromagnetism re-emerges in the superconducting
state in the presence of an Abrikosov vortex lattice
\cite{Kang03}, suggesting that the superconductivity competes with
a conventional antiferromagnet.

Different from the p-type systems, the n-type systems stay
insulating up to quite high dopings while the antiferromagnetism
of half filling degrades quite slowly. In fact, Vajk {\it et al.}
\cite{Vajk02} showed that the behavior of this antiferromagnet can
be understood in detail assuming a random dilution of the
Heisenberg quantum antiferromagnet, suggesting that the individual
carriers stay strongly bound to impurity sites. Is the field
induced antiferromagnet of a similar kind? Using NMR we will
present here evidences that in between the superconductor and the
strong pinning phase there is yet another phase. By reoxygenating
a superconducting ${\rm Nd_{2-x} Ce_{x} CuO_4}$ (NCCO) sample with
$x = 0.15$ up to the point that superconducting diamagnetism has
disappeared we have detected a phase in a very small oxygen-doping
range located in between the superconductor and the site diluted
antiferromagnet. This phase is extremely sensitive to further
oxygenation. Approximately one oxygen atom per 100 \cite{Onose99}
unit cells suffices to destroy this phase allowing a site-dilution
antiferromagnet to take over. With NMR one can measure the {\em
amplitude} distribution of the magnetic moments. These show a
double-peak distribution in this special phase. Extensive
Monte-Carlo simulations on a representative model show that such a
distribution will only arise in either a stripe phase or in a
situation where {\em large} droplets are formed. Since the neutron
scattering appear to rule out the former we conclude that the
superconductivity in the n-type systems competes with a genuine
type I Mott-insulating state \cite{Lee03}.

{\it Experimental.} The experiments were performed on single
crystals with $x$ = 0.15, 0.12 and 0.08. For $x=0.15$ the oxygen
reduced sample (NCCOp) had a superconducting transition
temperature of $T_{c}=21$~K. Half of the crystal (2x1x0.2 mm) was
reoxygenated in air at 900$^0$~C till no superconductivity could
be detected in the SQUID-susceptibility measurement (sample
NCCO1). After performing the full set of NMR measurements, the
same sample was oxygenated further under 4 bar of oxygen at
850$^0$~C for 20 hours (sample NCCO2), and NMR-scanned again.
Sample preparation and characterization are described elsewhere
\cite{Nugroho99}. The Cu NMR spectra were obtained with
conventional phase coherent pulsed NMR between 4 and 350~K by
sweeping the magnetic field at various constant frequencies. The
NMR line-shift was determined by using a gyromagnetic ratio of
1.1285~MHz/kOe for $^{63}$Cu and $K=0.238$~\% of metallic Cu as a
reference.
\begin{figure}[htb]
\begin{center}
\includegraphics[width=7.5cm]{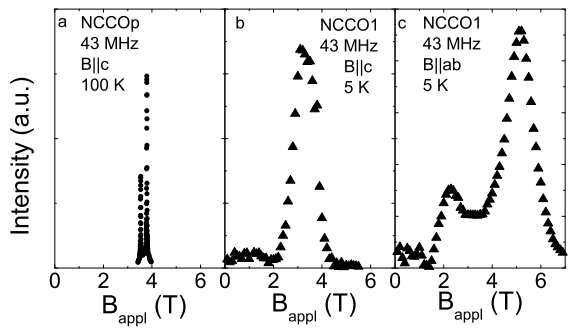}
\end{center}
\caption{Cu NMR spectra for $x=0.15$ in NCCOp (for comparison scaled
down from 78 to 43 MHz) at 100~K (at this temperature the spectra
for NCCO1 and NCCO2 are identical to that of NCCOp) and NCCO1 at
5~K. The broadening and shifts at low temperature are clearly
visible.} \label{spectra}
\end{figure}

The Cu NMR spectra were studied for two orientations ($B_{\rm
appl}||c$ and $B_{\rm appl}||ab$) of the crystals with respect to
applied external magnetic field $B_{\rm appl}$.  An example for
$x=0.15$ with $B_{\rm appl}||c$ is shown in Fig.~\ref{spectra}a at
high temperature (100~K). The two narrow lines correspond to the
central transitions of the two copper isotopes $^{63}$Cu and
$^{65}$Cu and are superposed on a relatively broad background
which likely originates from quadrupole satellites with small
quadrupole frequencies. For characterization of the samples, we
measured the $T$ dependencies of the normalized product of the
line intensity ($I$)(corrected for $T_2$ effects) and $T$, the NMR
line-sift $^{63}K$, and the fundamental relaxation probability
$W_1$ \cite{Curro00,AbuShiekah01,Hunt01,details}. Here we
concentrate on the low temperature field profiles on the Cu-sites
only - the internal field effects are so large that the role of
the Nd-moments can be neglected.

\begin{figure}[htb]
\begin{center}
 \includegraphics[width=7.5cm]{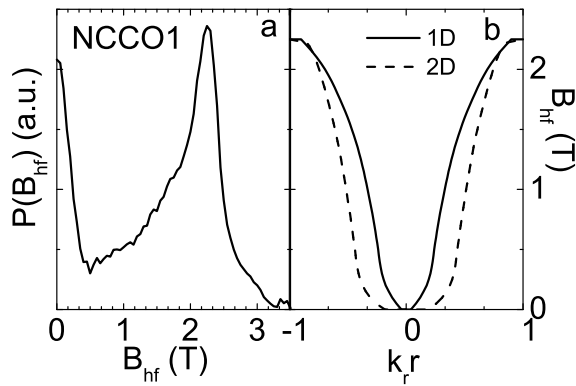}
\end{center}
\caption{(a) Amplitude distribution of the internal field for
NCCO1 at 5 K. The solid line gives the profile of the hyperfine
field, as deduced from $B \parallel c$. From the data for $B\perp
c$ the antiferromagnetic alignment of the electron spins in the
$(a,b)$-plane appears to be strongly disordered, see text. (b)
Simple reconstruction of the internal field distribution for a one
and two dimensional spin structure. In 1D the radial coordinate
$r$ with (incommensurate) wavevector $k_r$ have to be read as $x$
and $k_x$. In 2 D cylindrical symmetry is assumed.} \label{SDW}
\end{figure}

{\it Analysis of the low temperature profiles}. In the presence of
the static hyperfine magnetic field $B_{\rm hf}$ on the copper
nuclei produced by ordered Cu$^{2+}$ spins, one should observe two
Cu NMR lines at $B=[B_{\rm appl}^{2}+B_{\rm hf}^{2} + 2B_{\rm
appl}B_{\rm hf}\cos \alpha]^{0.5}$, where $\alpha$ is the angle
between the directions of $B_{\rm appl}$ and $B_{\rm hf}$. It is
known from neutron scattering experiments on ${\rm
Nd_{1.85}Ce_{0.15}CuO_{4}}$ \cite{Yamada99} that the Cu$^{2+}$
spins are ordered antiferromagnetically in the $ab$-plane.
Therefore, for $B_{\rm appl}||c$ ($\alpha=90^0$) only one Cu NMR
resonance is expected at a lower value than the field calculated
from the gyromagnetic ratio, $B_{0}$. For those nuclei, where for
some reason the internal field is cancelled, the resonance will be
unshifted. The Cu NMR spectra recorded for $B_{\rm appl}||c$ for
$x=$ 0.08, 0.13 and 0.15 can indeed be qualitatively explained in
this way: for $x=0.15$, see Fig.~\ref{spectra}b, part of the
nuclei mainly feel the external field (resonate at the fields of
Fig.~\ref{spectra}a) and the remainder experiences an internal
field in addition. At lower doping the line shifts are larger. The
spectra for $B_{\rm appl}||ab$, see Fig.~\ref{spectra}c, seem to
be composed out of at least two lines of different intensities.
This difference is not due to $T_2$ or different
$\pi/2$-conditions, because these are similar for both peaks
($T_{2}^{-1}=23(1)\times 10^{3}$~s$^{-1}$
and $\pi /2=2~\mu$s). \\

For a quantitative analysis, the low-temperature copper NMR
spectra  of NCCO1 at 5~K were simulated by exact diagonalization
of the nuclear spin Hamiltonian ${\cal H}$ with effective spin
$I=1/2$ (the quadrupolar interaction is weak compared to the
magnetic interaction) for both isotopes $^{63}$Cu and $^{65}$Cu.
For $B\parallel c$,  ${\cal H}$ can be written as ${\cal H}=\gamma
\hbar B_{\rm appl}I_{z}+\gamma \hbar B_{\rm hf}[I_{x} \cos \varphi
+I_{y}\sin \varphi ]+ \gamma \hbar B_{1}I_{x}$, where the
$z$-direction is along the crystallographic $c$-axis and $\varphi$
is angle between $B_{\rm hf}$ and the rf-field $B_{1}$ (along the
$x$-axis); for $B\perp c$, $I_z$ has to be replaced by $I_y$.
$B_{\rm appl}\perp B_{1}$ in all experiments and the AF coupled
electron spins are assumed to be always in the ($a,b$) plane. The
amplitude distribution of the internal field $P(B_{\rm hf})$ was
obtained via Monte-Carlo optimization to reproduce the data for
$B\parallel c$, see Fig.~\ref{SDW}a. Analysis of the $B\perp c$
data (Fig.~\ref{spectra}c) with this pure in-plane AF spin model
indicates that the alignment of the electron spins is rather
disordered in the ($a,b$)-plane, and on average show a field
induced canting of some 20 degrees, which does not influence
significantly the distribution $P(B_{\rm hf})$ in Fig.~\ref{SDW}a.
In the $x=0.12,0.08$ samples the computations give a single peak
at a field $\sim 3$~T, consistent with a simple commensurate
antiferromagnet. However, in the slightly oxygenated $x=0.15$
sample this spin-amplitude distribution is quite structured: the
peak at 2.4~T (corresponding with a reduced moment of $\sim 0.14
\mu_B$) is asymmetric showing a slow decrease on the smaller
moment side, while a second peak at vanishing internal field
corresponds with sites where the internal fields have completely
disappeared.

\begin{figure}[htb]
\begin{center}
\includegraphics[width=7.5cm]{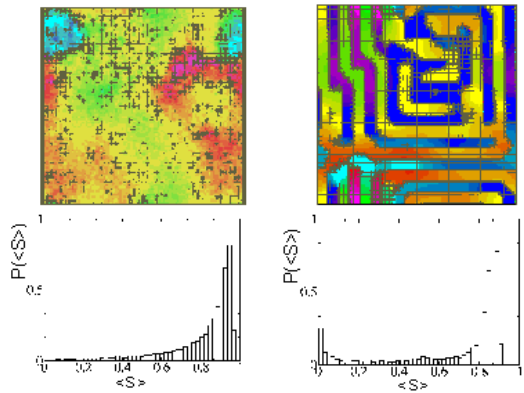}
\end{center}
\caption{ 2D map of the blob (a) and stripe (b)
 results from thermalized averaged MC simulations, obtained with the
 hamiltonian discussed in the text. The spin-orientational disorder
 in the $(a,b)$-plane is represented by a difference in color
 (complementary colors correspond to opposite aligned spins), while
 the color intensity corresponds to the spin value. The spin-zero
 regions (black), as also seen in experiment, are found in the
 stripe simulations, but are almost absent for small blobs. The
 parameters for the blob and  stripe panels are ($J$ = 1, $V_C$ =
 0): $J_1 = 0.2$, $Q = 0.2$, $V_S = 0.2$, $R = 10$, $T = 0.4$ and
 $J_1 = 1$, $Q = 0.4$, $V_S =  0.2$, $R = 10$ with $T = 0.3$.} \label{MCt}
\end{figure}

\begin{figure}[htb]
\begin{center}
\includegraphics[width=7.5cm]{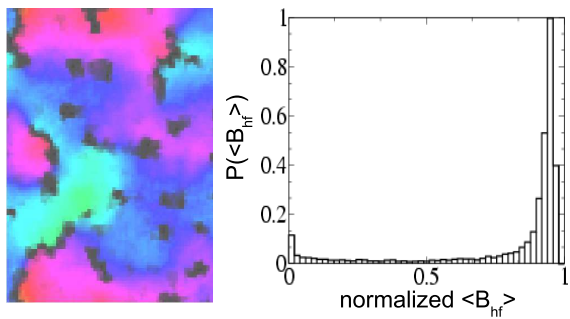}
\end{center}
\caption{Blob results from the MC simulations - a thermalized
result for larger blobs than in the previous figure. Only if blobs
are sufficiently large, zero field sites can be generated in the
presence of the transferred hyperfine interaction. For still
larger blobs the zero field contribution will further increase.
Parameter values are $J$ = 1, $T$ = 0.3, $J_1$ = 0, $Q$ = 0.04,
$V_S$ = 0.2 with $R$ = 10 and $V_C$ = - 1.5.} \label{MCf}
\end{figure}

{\em Interpretation.} If the internal-field distribution would
arise from a stripe-like spin structure with wave vector $k_x$, we
would have $P(B)={d(k_xx)}/{dB}$ and hence $x \propto
\int_{0}^{B/B_{\max }}P(B)dB$. Using the experimental outcome for
$P(B_{\rm hf})$ the integration leads to the 1D-wave depicted in
Fig.~\ref{SDW}b; note that because the Cu-nuclei that experience
$B_{\rm hf}$ have regular lattice positions, the wave vector needs
to be incommensurate to realize such a field distribution at the
Cu-sites. The peak of $P(B_{\rm hf})$ at zero internal field in
Fig.~\ref{SDW}a would then correspond with the nodes. However, the
field profile for a radially symmetric two dimensional spin
profile can be generated similarly via $(k_rr)^2 \propto
\int_{0}^{B/B_{\max }}P(B)dB$ and is given in the same figure.
These simple calculations are only meant to show that both stripe
like structures (comparable to the 1D case) and charge blobs (2D
case) are compatible with the experimental data. Furthermore the
needed zero field region in 2D is larger than in 1D (due to the
transferred hyperfine interaction the size of the blobs with zero
spin exceeds the zero field region).

The size enters into the problem even stronger because of the
specific form of the hamiltonian. The internal field has not only
a on-site contribution $A$, but also the transferred hyperfine
interaction $B \approx A/4$, which gives a field of opposite sign
and hence in a droplet a non-zero field will be generated on
zero-spin sites (in a stripe configuration zero field states are
easier to generate because it is an anti-domain boundary). To make
this case more persuasive we performed Monte-Carlo simulations on
a classical model designed to interpolate smoothly between
stripe-like and droplet-like ground state textures
\cite{Pijnenburg96}. This is a classical, spin-full lattice gas
model which is similar to the model introduced by Stojkovic {\em
et al.} \cite{Stojkovic99}. It builds in the quantum-physical
ingredient that electrons/holes (`charge', $\sim ( 1 - n_i)$)
cause anti-phase boundaries in the spin system by a `charge
mediated' exchange interaction \cite{Zaanen98} coupling the spins
antiferromagnetically across the charge, $H_{spin} = J \sum_{x,y}
S_{x,y} \cdot (S_{x,y+1} + S_{x+1,y}) + J_1 \sum_{x,y} S_{x,y}
\cdot [(1 - n_{x,y+1})S_{x,y+2} + (1 - n_{x+1,y}) S_{x+2,y}]$ with
$J, J_1 \geq 0$. The frustration in the spin system is released by
the formation of charged domain walls. The frustrated phase
separation motive is build in by balancing the spin-mediated
attractive charge-charge interactions with a repulsive $1/r$
Coulomb interaction of strength $Q$ between the charges, which we
cutoff at 10 sites (beyond the inter-stripe distance) for reasons
of numerical efficiency \cite{notecutoff}. We take a $XY$ spin
system and the spin-orientational disorder is build in by  a spin
pinning potential, breaking  the ground-state degeneracy by
choosing a direction (up to spin reflection) for the
antiferromagnetic background, as well as providing an
inhomogeneous charge potential. It has the form $H_{\rm{Pot}} =
-V_S|\sum_{x,y} S_{x,y} \cdot P_{x,y}|$, where $P_{x,y} =
\sum_{x',y'} A_{x',y'} e^{i \Theta_{x',y'}} \exp(-\sqrt{(x-x')^2 +
(y-y')^2} / R)$ with $R$ the correlation length over which the
antiferromagnetic background is allowed to rotate. The angle
$\Theta_{x',y'}$ is chosen randomly with a uniform distribution
and amplitude $A_{x',y'} = {\rm Random}[0,1]$. Finally we also
incorporate a quenched disorder potential acting on the charge in
the form of $V_C = {\rm Random}[0,1]$

Despite its simplicity this simple model is capable of generating
quite complex ground states, ranging from highly disordered
droplet patterns ($J_1 \rightarrow 0$) to very orderly stripe
patterns $J_1, J >> Q, V_S$. In the simulations we used small but
finite temperatures and thermal annealing, mimicking the effects
of quantum fluctuations in smearing the sharp textures
\cite{Zaanen01} in this classical model. Two typical outcomes are
illustrated in Figs.~\ref{MCt},\ref{MCf}. We have extensively
scanned the parameter space of this model. Invariably, we find
that a double peak structure is easily associated with stripe like
patterns (Fig.~\ref{MCt}b), but these configurations invariably
lead to incommensurate scattering amplitudes, which are not
observed experimentally. When droplets are formed instead, the
zero-moment peak is always lacking for small droplets
(Fig.~\ref{MCt} and Fig.~\ref{MCf}). Only droplets of 2 nm or more
(25 sites or larger) mimic the peculiarities of the NMR spectra.
This adds confidence to our claim that the double peak structure
in the local moment distribution is the NMR fingerprint signalling
the presence of large charge blobs.

\begin{figure}[htb]
\begin{center}
\includegraphics[width=7.5cm]{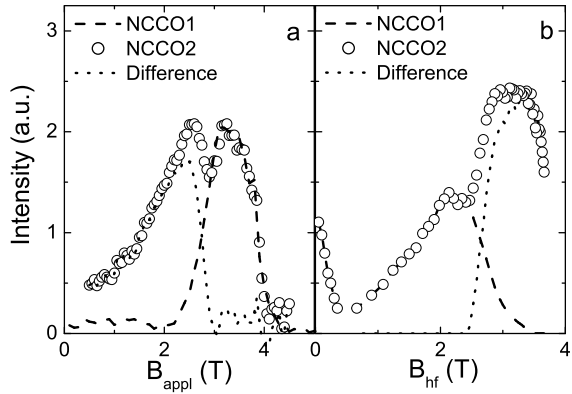}
\end{center}
\caption{Copper NMR spectra (the intensity in arbitrary units along
the vertical axis is plotted against the magnetic field value in
Tesla) at 5~K for NCCO1 and NCCO2 (a). The lines in (b) are fits
with an optimized internal field distribution, see text.}
\label{LTspec}
\end{figure}

How to reconcile our findings with the neutron scattering evidences
favoring the site-diluted antiferromagnet? The NMR sample has a
volume of 1 mm$^3$, much smaller than typical neutron targets and we
found the large droplet phase by almost continuously monitoring the
disappearance of the superconducting state while the sample was
oxygenated. Apparently, the large droplet phase in the NCCO1 sample
is linked to the presence of a tiny amount of  apical oxygens
($\delta \sim 0.005$ for NCCO1) \cite{Onose99}. The result for
continuing the oxygenation of NCCO1 in 4 bar of oxygen as used in
other studies \cite{Mang03} is given in Fig.~\ref{LTspec}. The data
show clearly that in NCCO2 the phase of NCCO1 is suppressed in favor
of a new phase \cite{noteNCCO2}. Applying a similar analysis as
before, we find that this new phase is characteristic of an
antiferromagnet with an internal field of 3~T - the phase seen in
the neutron data. Although it is hard to arrive at precise numbers,
it seems that one extra oxygen per roughly 100 unit cells suffices
to fracture the droplets and to favor a state where the charge carriers
are strongly bound to impurities.
As the superconductor, the large droplet insulating
phase shows an extreme sensitivity to oxygen concentration
explaining why this phase has been missed in the neutron scattering
studies.

In summary, we have presented evidence demonstrating that the
n-type cuprate superconductor is in a direct competition with a
phase where the charge carriers form large droplets in a
commensurate antiferromagnetic background. This phase is in turn
extremely sensitive to the chemical conditions. Tiny amounts of
excess oxygen suffice to destroy both superconductivity and the
large droplet phase, in favor of a phase where the charge carriers
are strongly bound to the impurities.

We acknowledge fruitful discussions with Martin Greven (Stanford)
and the financial support of FOM/NWO. Some of the numerical
calculations were performed at the APAC National Facility via a
grant from the Australian National University Supercomputer Time
Allocation Committee.


\begin{thebibliography}{10}
\bibitem{details}
{Details of the NMR relaxation and lineshift data will be
published elsewhere and are available on request.}
\bibitem{notecutoff} Using cut-offs larger than the typical length scale of the highly structured solutions
did not change the picture.
\bibitem{noteNCCO2}
{Prolongued oxydation is expected to turn the crystal completely
to the AF phase - we decided to keep the sample in its present
mixed state to allow further inspection.}
\end{thebibliography}
\end{document}